%% file: main.tex
\documentclass[runningheads]{llncs}
\usepackage[T1]{fontenc}
\usepackage{graphicx}

\usepackage[colorlinks,linkcolor=blue, citecolor=blue]{hyperref}
\usepackage{algpseudocode}
\usepackage[ruled,vlined,linesnumbered]{algorithm2e}
\usepackage{amsmath}
\usepackage{float}          
\usepackage{amsmath} 
\usepackage{bm}      
\usepackage{lineno}
\usepackage{graphicx}
\usepackage{subcaption}

\usepackage{color}

\begin{document}
\title{HAS-GPU: Efficient Hybrid Auto-scaling with Fine-grained GPU Allocation for SLO-aware Serverless Inferences}
\titlerunning{HAS-GPU}
%
%
\author{}
\institute{}

\author{Jianfeng Gu \inst{1} \and Puxuan Wang \inst{1} \and Isaac David Nunez Araya \inst{1} \and Kai Huang \inst{2} \and Michael Gerndt \inst{1}}
\institute{Technical University of Munich \and  Sun Yat-sen University \\
\email{jianfeng.gu@tum.de, puxuan.wang@tum.de, isaac.nunez@tum.de, huangk36@mail.sysu.edu.cn, gerndt@in.tum.de}
}

%


%
\maketitle              
\vspace{-3mm}
\begin{abstract}
Serverless Computing (FaaS) has become a popular paradigm for deep learning inference due to the ease of deployment and pay-per-use benefits. However, current serverless inference platforms encounter the coarse-grained and static GPU resource allocation problems during scaling, which leads to high costs and Service Level Objective (SLO) violations in fluctuating workloads. Meanwhile, current platforms only support horizontal scaling for GPU inferences, thus the cold start problem further exacerbates the problems. In this paper, we propose \textbf{HAS-GPU}, an efficient \textbf{H}ybrid \textbf{A}uto-scaling \textbf{S}erverless architecture with fine-grained \textbf{GPU} allocation for deep learning inferences. HAS-GPU proposes an agile scheduler capable of allocating GPU Streaming Multiprocessor (SM) partitions and time quotas with arbitrary granularity and enables significant vertical quota scalability at runtime. To resolve performance uncertainty introduced by massive fine-grained resource configuration spaces, we propose the Resource-aware Performance Predictor (RaPP). Furthermore, we present an adaptive hybrid auto-scaling algorithm with both horizontal and vertical scaling to ensure inference SLOs and minimize GPU costs. The experiments demonstrated that compared to the mainstream serverless inference platform, HAS-GPU reduces function costs by an average of \textbf{10.8x} with better SLO guarantees. Compared to state-of-the-art spatio-temporal GPU sharing serverless framework, HAS-GPU reduces function SLO violation by \textbf{4.8x} and cost by 1.72x on average.

\end{abstract}
\vspace{-4mm}
\section{Introduction}
\vspace{-2mm}
\input{intro}

\vspace{-5mm}
\section{Related Work}
\input{related_work}

\section{System Design}
\input{design}

\vspace{-2mm}
\section{Experiment and Evaluation}
\input{evaluation.tex}

\vspace{-4mm}
\section{Conclusion}
\vspace{-3mm}
\input{conclusion.tex}

\subsubsection{Artifact Availability} The artifact is available in the Zenodo repository \cite{gu2025artifact}. 

\subsubsection{\discintname} The authors have no competing interests to declare that are relevant to the content of this article.

%
%
\vspace{-4mm}
\bibliographystyle{splncs04}
\bibliography{ref}
%




\end{document}

%% file: intro.tex
Serverless computing, also referred to as Function-as-a-Service (FaaS), is emerging as a prominent paradigm for next-generation cloud-native computing due to the ease of deployment, high scalability, and cost-effective pay-per-use benefits. It shifts the burden of complex resource allocation and runtime maintenance from users to cloud providers, while its built-in agile scaling and event-driven policies enable applications to dynamically adapt to fluctuating workloads on demand, thereby reducing resource usage and the cost per request. Traditional FaaS platforms primarily support CPU functions, such as AWS Lambda \cite{aws_lambda} and Google Run Function \cite{google_run_function}. 
 However, with the growing prevalence of deep learning (DL) applications, a rising number of inference tasks are being deployed on GPU-enabled serverless computing platforms, such as Azure Functions \cite{azure_gpu_function}, Alibaba Cloud Function \cite{alibaba_cloud_function}, KServe \cite{kserve}, and RunPod \cite{runpod}.
 
Nevertheless, current serverless inference platforms commonly encounter problems with coarse GPU allocation and limited scalability. First, with advanced GPU manufacturing, modern GPUs integrate more compute units and memory resources in a single board, such as NVIDIA V100 (80 SM (Streaming Multiprocessor) units, 5120 CUDA cores) and H100 (144 SMs, 18432 CUDA cores). The rise of large language models (LLMs) has further driven rapid deployment of high-end, expensive GPUs in modern data centers. Unlike LLM inference, which requires exclusive access to multiple GPUs and customized systems, serverless inference platforms typically run smaller deep learning models \cite{alibaba_cloud_use_case}\cite{ali2022optimizing}\cite{yang2022infless} in multi-tenant environments. However, current GPU-based serverless platforms \cite{azure_gpu_function}\cite{kserve}\cite{runpod} simply allocate an entire GPU to a single function instance, even though most inference tasks fall far short of fully using the GPU resources. This coarse GPU resource allocation leads to low GPU utilization and increased function costs.

Second, some approaches attempt to enable multiple function instances to share a GPU, but simultaneously undergo the problems of significant scalability limitations and potentially frequent Service Level Objectives (SLOs) violations. Current spatial GPU sharing approaches, such as NVIDIA's Multi-Instance GPU (MIG) support in Kubernetes \cite{mig_gpu}, Alibaba Cloud's cGPU  \cite{alibaba_cgpu}, and the MPS-based method GSlice \cite{dhakal2020gslice}, enable the allocation of partial GPU compute units to applications. Meanwhile, other approaches \cite{choi2022serving}\cite{gu2023fast} introduce spatial and temporal resource allocations. However, these approaches can only statically allocate fixed-size GPU resources to inference tasks. When dealing with highly fluctuant serverless workloads, they can only rely on horizontal scaling, which incurs significant cold start overhead due to the creation of new instances, particularly for deep learning models that require massive model data loading. Unlike serverless CPU functions, which can flexibly scale vertically by adjusting CPU cores/quota and memory via cgroups system, there is currently a lack of system for achieving fine-grained vertical scaling on GPUs. The limitation in vertical scalability prevents GPU functions from effectively ensuring function SLOs.

In this paper, we proposed \textbf{HAS-GPU}, an efficient \textbf{H}ybrid \textbf{A}uto-scaling \textbf{S}erverless architecture with both vertical and horizontal scaling and fine-grained \textbf{GPU} allocation for deep learning inferences. HAS-GPU incorporates an agile scheduler capable of allocating GPU SM partitions and time quotas with arbitrary granularity and enables dynamic GPU quota reallocation at runtime. The flexible GPU temporal resource reallocation provides significant support for function vertical scaling. HAS-GPU can quickly respond to burst workloads by increasing the time quota and provide a time buffer for horizontal scaling. Meanwhile, it can optimize time quota allocation during low request periods and sustains a keep-alive state with minimal resource consumption, eliminating cold start overhead from scale-to-zero and significantly reducing function costs.

Furthermore, since finer-grained GPU resource allocation also implies a significantly larger search space, we propose an accurate \textbf{R}esource-\textbf{a}ware \textbf{P}erformance \textbf{P}rediction (\textbf{RaPP}) model to facilitate spatio-temporal GPU resource allocation. The model addresses the inference performance uncertainty introduced by massive resource configuration spaces. RaPP integrates and learns static and runtime features of deep learning operators and computing graphs under resource constraints, enabling accurate latency prediction for different batch sizes and models across any spatio-temporal GPU resource configurations. This eliminates the need for large-scale pre-profiling required in previous work \cite{choi2022serving}\cite{gu2023fast}\cite{yang2022infless}.

Meanwhile, to handle highly fluctuating serverless workloads, we propose an adaptive hybrid auto-scaling algorithm. The algorithm introduces the co-design of fine-grained GPU resource allocation and function scheduling. By efficiently coordinating vertical and horizontal scaling, it enables functions to dynamically and flexibly adjust GPU resources with a fine granularity at runtime to meet their SLOs. Meanwhile, the high elasticity minimizes unnecessary resource consumption, effectively reducing function costs. Moreover, the algorithm introduces SM partition alignment-based GPU resource allocation, effectively addressing resource fragmentation problem in fine-grained allocation.

In a nutshell, the contributions are summarized as follows:
\vspace{-2mm}
\begin{itemize}
\item We propose \textbf{HAS-GPU}, an efficient hybrid auto-scaling architecture with fine-grained GPU allocation for serverless inferences to effectively ensure function SLOs and reduce function cost. To our best knowledge, HAS-GPU is the first work providing GPU vertical scaling for serverless computing.

\item We propose \textbf{RaPP}, an accurate resource-aware performance prediction model, addressing the problem of massive pre-profiling requirement and inference performance uncertainty introduced by massive resource configuration spaces.

\item We propose the \textbf{hybrid auto-scaling algorithm} to facilitate agile vertical and horizontal scaling. The algorithm introduces the co-design of fine-grained GPU resource allocation and function scheduling, effectively ensures function SLO, reduces function cost, and avoids resource fragmentation.

\item We implement the HAS-GPU architecture from low-level GPU device management to high-level serverless function scheduling. Experiments on the MLPerf-based benchmark \cite{reddi2020mlperf} and Azure Trace workload \cite{zhang2021faster} demonstrate that, compared to mainstream serverless inference platforms, HAS-GPU reduces function costs by 10.8x on average with better SLO guarantees. Compared to the state-of-the-art spatio-temporal GPU sharing framework, HAS-GPU reduces function SLO violation by 4.8x and cost by 1.72x on average. 
\end{itemize}

%% file: related_work.tex
\vspace{-3mm}
\subsection{Serverless Inference}
\vspace{-2mm}
With the widespread adoption of deep learning (DL) applications, serverless computing, commonly known as Function-as-a-Service (FaaS) \cite{jonas2019cloud}, has become a popular choice for deploying DL inference applications \cite{alibaba_cloud_use_case}\cite{aslani2025machine}. FaaS platforms offer seamless scalability while abstracting away complex resource management. Additionally, their event-driven, pay-per-use pricing model helps reduce costs. Major cloud providers, including AWS SageMaker \cite{amazon_sagemaker}, Azure Functions \cite{azure_gpu_function}, and Alibaba Cloud \cite{alibaba_cloud_function}, have introduced serverless inference platforms. Meanwhile, various research efforts have introduced optimizations for serverless inference architectures, such as tensor sharing \cite{li2022tetris} and request batching \cite{ali2022optimizing}. However, most of these approaches perform inference on CPU-based functions, while GPU-based architectures \cite{azure_gpu_function} \cite{kserve} \cite{runpod} typically allocate an entire GPU to a single inference function instance even though the function cannot fully utilize it. As expensive high-end GPUs are increasingly deployed in cloud and data centers, the cost of function inference continues to rise. Therefore, achieving finer-grained GPU resource allocation for inference functions is critical to reducing inference costs.
\vspace{-4mm}
\subsection{Fine-grained GPU Allocation}
\vspace{-1.7mm}
With advancements in GPU manufacturing, modern GPUs integrate more SMs, CUDA cores, and memory on a single board, such as the V100 (80 SMs, 5120 CUDA cores) and H100 (144 SMs, 18,432 CUDA cores). Therefore, research has increasingly focused on finer-grained GPU allocation and sharing to minimize resource waste. To enable GPU sharing, NVIDIA introduced MIG \cite{mig_gpu} for hardware-based partitioning and MPS \cite{mps_gpu} for software-based isolation, while Alibaba Cloud implemented cGPU \cite{alibaba_cgpu} for GPU partitioning in the Linux kernel.  Building on these, studies \cite{cho2022sla}\cite{dhakal2020gslice} have proposed spatial resource allocation and optimization strategies to meet application SLOs. Additionally, some approaches \cite{choi2022serving}\cite{gu2023fast}\cite{han2024inss} explore spatio-temporal GPU resource allocation, leveraging workload-based resource management to reduce applications' mutual interference, enhance application throughput, and improve GPU utilization. However, these approaches can only statically allocate fixed-size GPU resources to inference tasks. 

\vspace{-4mm}
\subsection{Performance Prediction for Deep Learning Inferences}
\vspace{-1.7mm}
Inference performance prediction is a crucial technique for reducing the need for extensive pre-profiling. Previous work nn-Meter \cite{zhang2021nn} predicts DL model latency at the kernel level by detecting kernels and summing the latency predictions from per-kernel predictors. But the model is limited to edge devices. DIPPM \cite{panner2023dippm} and NNLQP\cite{liu2022nnlqp} utilize static model computation graphs and operator features with graph neural network learning to predict the latency, memory, or energy consumption. However, these methods cannot predict performance under different runtimes and fine-grained GPU resource configurations. Currently, there are no methods for predicting model performance across a wide range of different fine-grained GPU resource allocations.
\vspace{-4mm}
\subsection{Horizontal and Vertical Auto-scaling}
\vspace{-1.7mm}
The CPU-based horizontal and vertical scaling has been widely studied in serverless computing \cite{tari2024auto}. However, GPU function is primarily limited to horizontal scaling, and almost all cloud providers \cite{alibaba_cloud_function}\cite{azure_gpu_function}\cite{runpod} only offer horizontal scaling for GPU function. This is mainly because CPU and memory resources can be quickly expanded using the cgroups system, whereas no system currently exists to achieve fine-grained vertical scaling on GPUs. Choi et.al \cite{choi2022serving} and FaST-GShare \cite{gu2023fast} proposed to select and scale inference functions with the most efficient spatio-temporal GPU resource configuration to meet a workload. INFless \cite{yang2022infless} introduced a horizontal non-uniform scaling policy with heterogeneous CPU/GPU resource allocation to maximize resource eﬃciency. GSLICE \cite{dhakal2020gslice} horizontally replaced functions with different spatial GPU resource allocations using the shadow functions. When dealing with highly fluctuating serverless workloads, these methods essentially only rely on horizontal scaling, in which the creation of new instances suffers from significant cold start overhead. Therefore, introducing vertical scaling for GPU functions is essential to further ensure function SLOs.

%% file: design.tex
\vspace{-7mm}
\begin{figure}[htbp]
  \centering
  \includegraphics[width=0.95\textwidth]{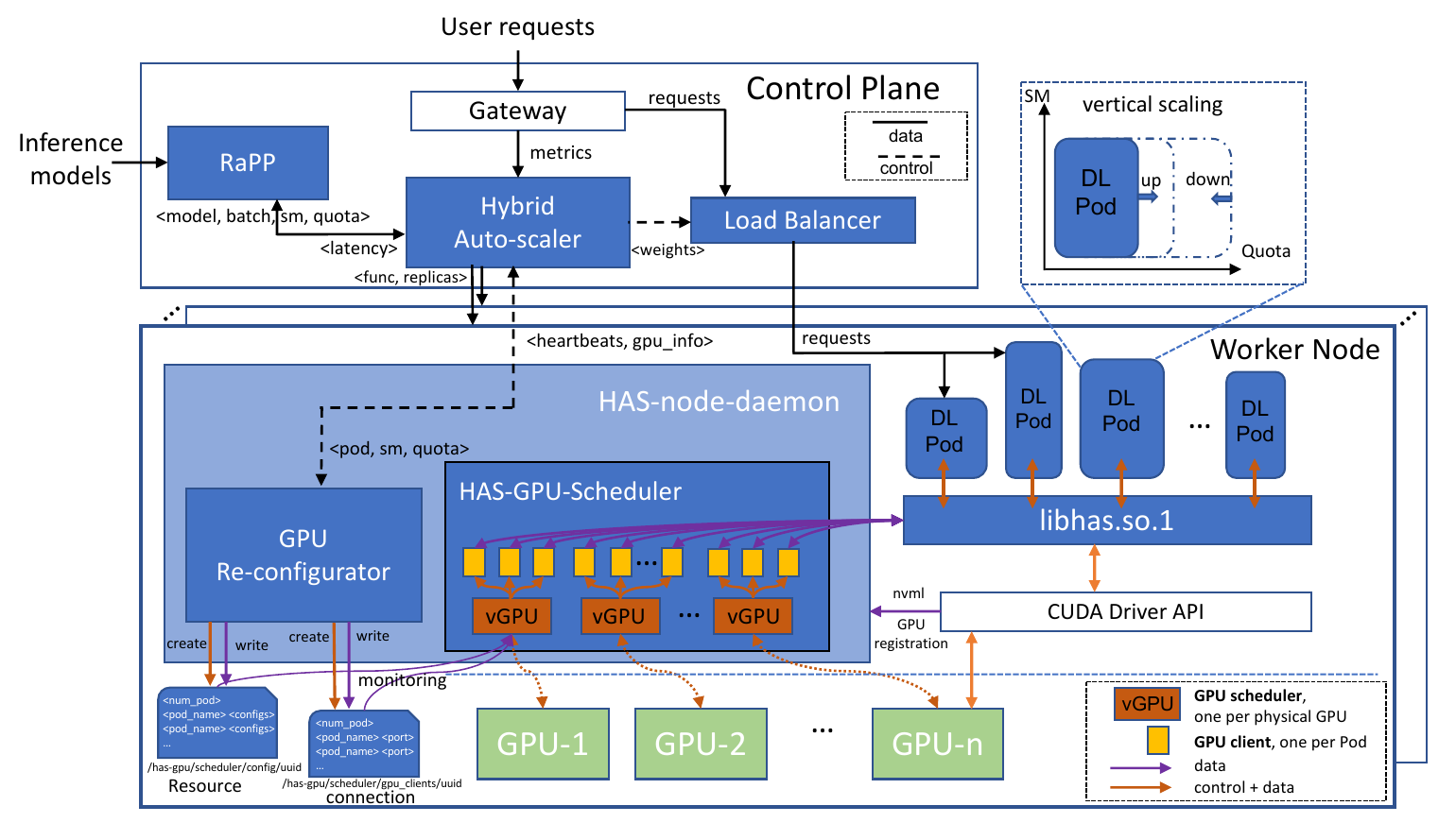}
  \vspace{-2mm}
  \caption{The architecture of HAS-GPU.}
  \label{fig:has_gpu_framework}
\end{figure}
\vspace{-6mm}
The design and workflow of HAS-GPU is shown in Figure \ref{fig:has_gpu_framework}. The architecture follows the fundamental structure of Kubernetes to support more FaaS platforms and mainly consists of five core components:  the Hybrid Auto-Scaler and performance prediction modules (RaPP) on the control plane, as well as the GPU Re-configurator, HAS-GPU-Scheduler, and \textit{libhas} modules on the worker nodes.

In the control plane, when a developer submits a DL inference model, \textbf{RaPP} automatically extracts the model and runtime features for performance analysis. By predicting latency and corresponding throughput capability under different batch sizes and fine-grained resource allocations, RaPP provides precise function performance information for the Hybrid Auto-Scaler. The \textbf{Hybrid Auto-Scaler} maintains at least one instance with minimal resources for each DL function and continuously retrieves request metrics from the Gateway. When significant fluctuations in user requests occur, the auto-scaler evaluates current pod instances of the function and GPU resource usage in the cluster. Based on the hybrid auto-scaling algorithm, the auto-scaler decides to apply either horizontal or vertical scaling and perform resource allocation and pod scheduling for the function. Meanwhile, the load balancer is updated with request distribution information according to the throughput capability of different function pods.

In GPU worker nodes, \textbf{HAS-node-daemon} manages the resource allocation and scheduling of all GPUs within a node and runs on each node. Inside HAS-node-Daemon, \textbf{HAS-GPU-Scheduler} abstracts each physical GPU device into a vGPU and creates a GPU client for each assigned pod to manage its GPU resource usage. Each vGPU coordinates and controls the GPU usage of its assigned GPU clients at runtime and is associated with two resource configuration device files in the host system. The \textbf{GPU Re-configurator} dynamically monitors the status of all GPUs within the node and provides real-time GPU information to the Hybrid Auto-Scaler. Meanwhile, it receives fine-grained GPU resource allocation instructions from the auto-scaler for pods and writes this information to the device files. The \textbf{\textit{libhas}} serves as the unified interface for resource control of pods in HAS-GPU. At runtime, pods utilize this library to request and obtain GPU resources from the corresponding GPU client for execution.
\vspace{-4mm}
\subsection{Fine-grained GPU Resource Allocation and Reallocation}
\vspace{-2mm}
HAS-GPU enables fine-grained allocation of GPU resources through spatio-temporal resource isolation and sharing. This is achieved by leveraging CUDA Driver API interception and MPS-based \cite{mps_gpu} Streaming Multiprocessor (SM) partitioning techniques. Currently, the proprietary ecosystem of GPU software stacks like CUDA makes it difficult for the system to control the execution prioritization and scheduling of DL tasks when multiple tasks are running together. 

\begin{figure}[htbp]
  \centering
  \begin{subfigure}[b]{0.35\textwidth}
    \includegraphics[width=\textwidth]{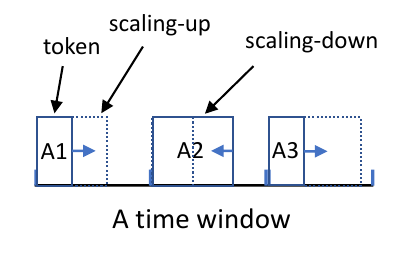}
    \label{fig:time_window}
  \end{subfigure}
  \begin{subfigure}[b]{0.52\textwidth}
    \includegraphics[width=\textwidth]{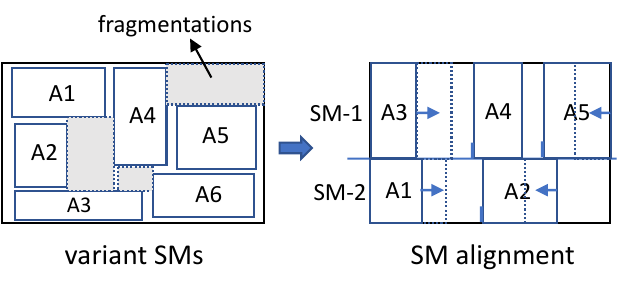}
    \label{fig:sm_alignment}
  \end{subfigure}
  \vspace{-5mm}
  \caption{Flexible vertical scaling and SM alignments to avoid fragmentations.}
  \label{fig:temporal_scaling}
  \vspace{-6mm}
\end{figure}

However, all deep learning inference tasks ultimately invoke the underlying unified CUDA Driver APIs, such as allocating GPU memory through \textit{cuMemAlloc()}, transferring data from host memory to device memory via \textit{cuMemcpyHtoD()}, and launching kernels using the \textit{cuLaunchKernel()} function. Therefore, HAS-GPU introduces new custom unified APIs and leverages the function interposition technique to load the new shared library \textit{libhas}. The library overrides functions in the standard CUDA library, seamlessly and effectively intercepting the CUDA function calls at runtime. Within the intercepting APIs, we design our resource allocation and scheduling mechanisms. We leverage the intercepted functions related to GPU memory allocation and release to enforce limits on the available GPU memory within a pod. As shown in Figure \ref{fig:has_gpu_framework}, the communication between the pod and its GPU client in the HAS-GPU-Scheduler is established through the intercepted \textit{cuLaunchKernel()} function. A pod must request a time token from the vGPU via the GPU client to execute CUDA kernels. By specifying the proportion of time tokens allocated within the time window of a vGPU, HAS-GPU achieves any fine-grained temporal GPU resource allocation for pods. When a pod requires vertical scaling, the HAS-GPU-Scheduler can dynamically modify the time token within the time window to achieve GPU resource reallocation with minimal overhead, as shown in Figure \ref{fig:temporal_scaling}. Meanwhile, the time window can be flexibly adjusted, similar to the CPU subsystem in cgroups, to accommodate varying temporal granularity requirements.

For spatial resource allocation, NVIDIA offers the Multi-Process Service (MPS) \cite{mps_gpu} interface, enabling systems to allocate arbitrary proportions of Streaming Multiprocessors (SMs) to certain applications. However, the allocated SMs are deeply tied to the CUDA context and must be specified when the context is initially created, preventing dynamic reallocation of SM resources at runtime. But dynamic SM allocation can easily lead to severe resource fragmentation, as shown in Figure \ref{fig:temporal_scaling}. Therefore, HAS-GPU achieves vertical scaling for a pod by leveraging flexible temporal resource reallocation under a stable SM allocation and by performance prediction across varying configurations. Meanwhile, a pod can be initially assigned any SM partitions if the GPU has no prior allocation.

Traditional serverless inference platforms \cite{kserve} manage GPU resources through the Kubernetes device plugin. However, the plugin only allows allocating GPUs at the instance level and cannot specify particular GPUs, thus hindering fine-grained GPU resource allocation. As illustrated in Figure \ref{fig:has_gpu_framework}, GPU Re-configurator bypasses the device plugin by directly managing GPU topology via NVML and uniquely identifying GPUs through their UUIDs. This enables the auto-scaler to accurately schedule pods to specific nodes and GPUs, and to update the pod connection and resource reconfiguration information to the specific device files.

\vspace{-2mm}
\subsection{Resource-aware Performance Prediction (RaPP)}
\vspace{-8mm}
\begin{figure}[htbp]
  \centering
  \includegraphics[width=0.95\textwidth]{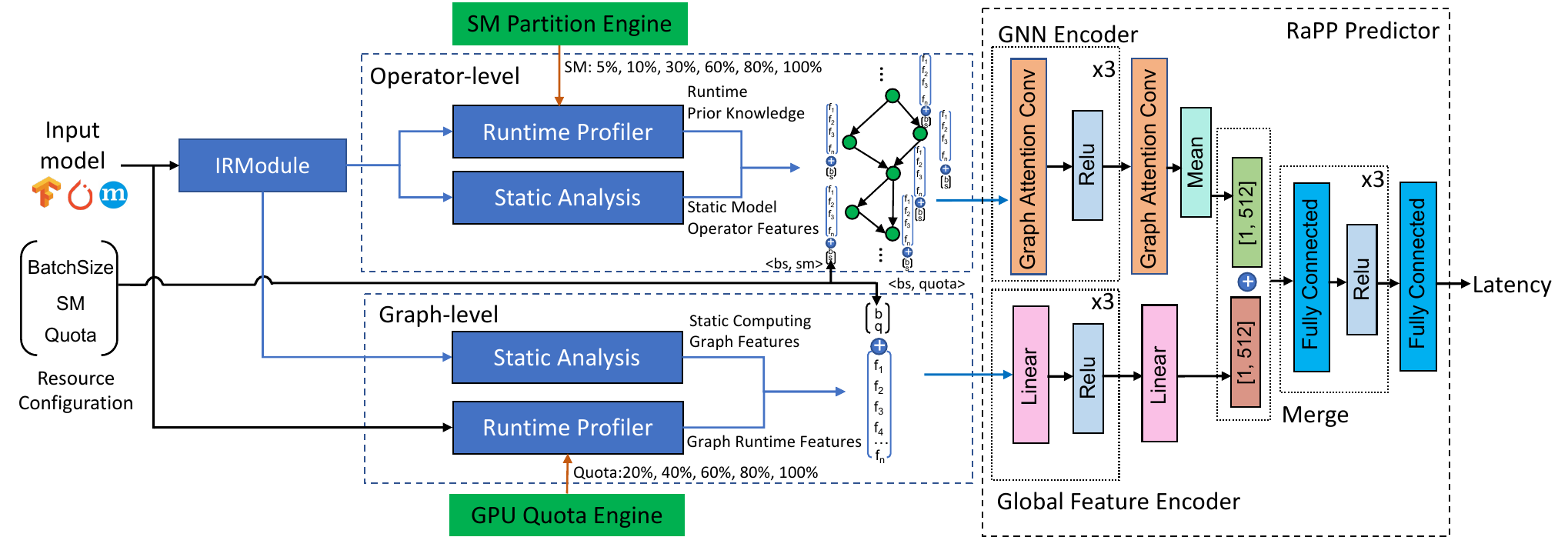}
  \caption{The Resource-aware Performance Prediction Model.}
  \label{fig:has_rapp}
  \vspace{-6mm}
\end{figure}
HAS-GPU supports resource allocation with any granularity but also introduces massive configuration spaces and performance uncertainty. For example, a deep learning model with 4 batch sizes, 10 quotas, and 10 SM partition configurations brings 400 distinct configuration possibilities and performance outcomes. Relying on traditional pre-profiling methods would incur significant resource and cost consumption. Therefore, HAS-GPU introduces the Resource-aware Performance Prediction (RaPP) model to predict latency for arbitrary batch sizes under any spatio-temporal GPU resource configurations. As shown in Figure \ref{fig:has_rapp}, RaPP comprises two main components: feature extraction and the GNN-based predictor. Traditional static feature-based methods \cite{panner2023dippm}\cite{liu2022nnlqp} are not suitable for resource-aware latency prediction, as different SMs and time quotas can significantly affect the execution time of operators and the overall computational graph at runtime. RaPP integrates both static and runtime characteristics at the operator and graph levels to provide a more accurate feature representation.


For operator-level feature extraction, RaPP first transforms the model into TVM's Relay IRModule \cite{chen2018tvm}. Relay IRModule serves as a unified intermediate representation for computational graphs and operators, offering compatibility across various DL frameworks. RaPP introduces an operator-level Runtime Profiler based on TVM's \textit{debug\_executor} and enables the collection of runtime statistics for each individual operator under various SM partition configurations. The Runtime Profiler perform operator profiling under a full time quota and six distinct SM configurations. This profiling data is subsequently incorporated as runtime prior knowledge into the operator feature representation graph. We adopt a full time quota since time-window-based quota allocation affects only the latency of the overall computational graph, without impacting the performance of individual operators. The SM configuration is not limited to six types and can be adjusted based on the complexity of the model operators. The execution time of operators also directly integrates performance information related to different GPU architectures and their SM characteristics. Meanwhile, similar to prior studies \cite{panner2023dippm}\cite{liu2022nnlqp}, we further incorporate static operator features to the feature graph, such as operator type, kernel size, channel, stride, and so forth.

For graph-level feature extraction, RaPP also incorporates static analysis and a runtime profiler. Specifically, RaPP leverages IRModule to collect static graph features, such as the number of floating-point operations, multiply-accumulate operations, and counts of key operators like \textit{nn.conv} and \textit{nn.dense}, since these features reflect data transfer overhead between GPU and host and the total computational load. Meanwhile, the runtime profiler evaluates the model under a full SM configuration and five distinct quota configurations to obtain the distribution of quota impacts on model computation.

RaPP integrates the input of batch size, quota, and SM partition configuration into operator and graph features for learning. Inside the RaPP Predictor, we utilize multiple Graph Attention Convolution (GAT) \cite{velivckovic2017graph} blocks to encode the operator feature graph. The attention mechanism in GAT helps to capture potential kernel fusion information among neighboring operators. Meanwhile, the predictor utilizes an MLP to encode the global features. By merging and learning from two types of features, the predictor estimates model inference latency.

\vspace{-3mm}
\subsection{Hybrid Auto-scaling}
\vspace{-1.5mm}
This section presents the hybrid auto-scaling algorithm which utilizes the cooperation of vertical and horizontal scaling to ensure function SLOs under fluctuating workload. The algorithm introduces the co-design of fine-grained GPU resource allocation, function scheduling, and GPU cluster resource management. 

Traditional auto-scaling methods \cite{tari2024auto} typically integrate workload prediction within the algorithm design. However, workloads across various scenarios often exhibit distinct distribution, making it difficult to develop a universal prediction model. To address this challenge, the HAS autoscaler decouples the request prediction model from the auto-scaling algorithm, enabling integration with alternative prediction models. For fluctuating serverless workloads, HAS-GPU proposes a Kalman filter-based short-term estimation approach that predicts the next request workload $R$ by the current measured request load $R_{t}$. The $R_{t}^{'}$ and $P_{t}^{'}$ represent the predicted workload and covariance based on the previous prediction. The $K$ is the Kalman gain, balancing the weights of the predicted and observed request workload in the final estimate. By integrating predictions with observations, the request predictor can efficiently adapt to fluctuating workloads.

\vspace{-2mm}
{\small{
$$
R_{t}^{'} = AR_{t-1} \, , \quad P_{t}^{'} = AR_{t-1}A^{T} + Q
\label{algo:rps_predict}
$$
\vspace{-6mm}
$$
K = \frac{P_{t}^{'}H}{HP_{t}^{'}H^{T} + D}, \quad R = R_{t}^{'} + K(R_t - HR_{t}^{'}), \quad P = (1-KH)P_{t}^{'}
\label{algo:rps_update}
$$
}}

As shown in Algorithm \ref{alg:hybrid_auto_scaling_algorithm}, once obtaining the predicted RPS (requests per second) $R$ of a function $f$, the hybrid auto-scaling algorithm starts to perform auto-scaling based on the function's existing pods and their GPU resource usage $P_f$, as well as current GPUs' occupancy $\{G_j\}$ across the cluster $G$. The auto-scaler determines the total processing capability of the function's currently running pods at first (Line 1). When the predicted RPS reaches the processing capability threshold $\alpha$, it triggers scaling up. This threshold helps prevent frequent scaling operations. Additionally, users can adjust it based on their desired sensitivity to scaling up and the required redundancy. To fill the RPS gap $\Delta R$, the auto-scaler tries vertical scaling by adding more quotas to pods at first (Lines 3 - 9). Pods with larger SM partitions are prioritized, as a smaller quota increase can provide a greater boost in throughput capability. For a scaling pod, the system first determines its maximum expandable quota based on its located GPU and SM partition type,  then incrementally increases the quota by step size $\Delta I_q$ to match the required RPS gap. Pods within a GPU are managed using SM alignment to prevent resource fragmentation, as shown in Figure \ref{fig:temporal_scaling}. New pods must either follow existing SM configurations or introduce new SM types without exceeding this limit. If the RPS gap remains unmet after vertical scaling, the system selects the least utilized GPU among the used GPUs for horizontal scaling (Lines 10 - 17). We define a new metric, HAS GPU Occupancy (HGO), to evaluate GPU utilization. If no GPU has sufficient resources to meet the RPS gap, the pod is deployed on a new GPU (Lines 18 - 19).

When the predicted RPS falls below a certain threshold $\beta$ of the pods' processing capacity, the auto-scaler triggers the scaling-down (Lines 20 - 26). To prevent frequent scaling,  a minimum interval $T_{cooldown}$ is enforced between consecutive scale-down operations, and at least one pod should be retained to guarantee a minimum request capacity $R_{min}$ and avoid the cold start. Pods with smaller SM partitions are prioritized to vertical scaling-down to guarantee potential processing capability. The auto-scaler follows the same stepwise vertical scaling-down and horizontal scaling-down as scaling-up.

\input{algorithms/algo_autoscaling}

%% file: algorithms/algo_autoscaling.tex
\begin{algorithm}[H]
\caption{\footnotesize{Hybrid Vertical and Horizontal Auto-Scaling.}}
\small
\label{alg:hybrid_auto_scaling_algorithm}
\SetKwInput{KwIn}{Input}
\SetKwComment{Comment}{/* }{ */}
\SetKwIF{If}{ElseIf}{Else}{if}{then}{else if}{else}{end if}
\SetKwFor{While}{while}{do}{end while}

\KwIn{
    $\bm{f}$: Inference function; \, $\bm{P_f = \{P_i\}}$: pod instances $P_i$ of function $f$; \\
    $\bm{R}$: Predicted RPS of the function; \, $\bm{G = \{G_j\}}$: The GPU $G_j$ in the cluster $G$;
}
\KwOut{ $\bm{{S_f} = \{S_i\}}$: Scaling actions $S_i$ for function $f$; \ $S_i = (f, P_{i}^{'}, \text{type})$;
} 
%
$C_f = \sum{C_{P_i}}, {P_i} \in {P_f}$, where $C_{P_i} = \text{RaPP}(f, b_i, s_i, q_i)$; // {\scriptsize{current processing capability.}} \\
{\scriptsize\tcp{Scaling Up.}} 
\If{$ R  > C_f * \alpha$}{

    $\Delta R = R - C_f * \alpha $; \quad $P^{'}_{f} = \text{sort}_{\downarrow {s_i}}(\{P_i\}), P_i \in P_f$; \, // {\scriptsize{Pods with more SMs first.}} \\
    
    {\scriptsize\tcp{Try vertical scaling-up first by adding more quota to pods.}} 
    \ForEach{$P_i \in P^{'}_{f}$ \text{and} $\Delta R > 0$}{
        $A_q = $ \text{RetriveMaxAvailQuotaForPod}($P_i$, $G_j$), \, $P_i$ runs in $G_j$; $\Delta C^{'} = 0$; $n = 1$; \\
        \While{$q_i +  \Delta {I_q} \times n \leq A_q$ and $\Delta R - \Delta C^{'} > 0$}{
            $C_{P_i}^{'}$ = \text{RaPP}($f$, $b_i$, $s_i$, $q_i + \Delta I_q \times n$); \, $\Delta C^{'} = C_{P_i}^{'} - C_{P_i}$; $n=n+1$;
        } 
        $P_i^{'} \gets (b_i, s_i, q_i + \Delta {I_q} \times n)$; \,  \bm{$S_i = (f, P_{i}^{'}, \bm{\rightarrow})$}; \, // vertical scale-up.  \\ $S = S \cap S_i$; \, $\Delta R = \Delta R - \Delta C^{'}$;
    }
    
    {\scriptsize\tcp{{\bfseries Horizontal scaling-up if vertical scaling is insufficient.}}} 
    \If{$\Delta R > 0 $}{
         {\scriptsize\tcp{{\bfseries Horizontal scaling to the used GPU with lowest HGO first.}}} 
        $G_{i} = \arg\min_{G_j}{\{H_{G_j}\}}$, where $H_{G_j} = \sum_{P_i}{s_i \times q_i}, \forall P_i$ run in $G_j$; \\  
        $(s_{max}, q_{max})$ = RetriveMaxAvailQuotaAndSM($G_{i}$); \\
        $C_{max}$=RaPP($f$, $s_{max}$, $q_{max}$) \\
        \If{$C_{max} > \Delta R$}{
            \While{$\Delta {I_q} \times n \leq q_{max}$ and $\Delta R - C_{P}^{'} > 0$}{
                $C_{P}^{'}$ = \text{RaPP}($f$, $b_i$, $s_i$, $\Delta I_q \times n$); \, $n=n+1$; \\
            } 
            $P^{'} \gets (b_i, s_i, \Delta {I_q} \times n)$; \,  \bm{$S_i = (f, P^{'}, \bm{\uparrow})$}; \, // new pod instance.
        }
    }
    {\scriptsize\tcp{{\bfseries Horizontal scale-up to a new GPU $G^{'}$ if used ones fall short.}}} 
    \If{$\Delta R > 0 $}{
        $(b^{'}, s^{'}, q^{'})$ = RaPPbyThroughput($f$, $\Delta R$); // Most efficient for $\Delta R$. 
        $P^{'} \gets (b^{'}, s^{'}, q^{'}, G^{'})$; \, $G=G \cap G^{'}$; \, \bm{$S_i = (f, P^{'}, \bm{\uparrow})$}; \, // new pod;
        
    }
}
{\scriptsize\tcp{Scaling Down.}} 
\If{$(R  < C_f \times \beta)$ \ and $(R > R_{min})$ and \ $(t > T_{cooldonw})$}{
    $\Delta R = C_f - R$; \quad $P^{'}_{f} = \text{sort}_{\uparrow {s_i}}(\{P_i\}), P_i \in P_f$; \, // {\scriptsize{fewer SMs first.}} \\
    Reduce the quota progressively in the same stepwise manner until $\Delta R \leq 0$;
    $P_i^{'} \gets (b_i, s_i, q_i - \Delta {I_q} \times n)$; \,  \bm{$S_i = (f, P_{i}^{'}, \bm{\leftarrow})$}; \, // vertical scale-down.  \\
    \If{$q_i - \Delta {I_q} \leq 0 $}{
        \bm{$S_i = (f, P_{i}^{'}, \bm{\downarrow})$}; \, // horizontal scale-down. 
    }
    $S = S \cap S_i$; \,
    \If{$\emptyset$ run in $G_j$}{
        $G = G \setminus G_{j}$;
    }
    
}
\end{algorithm}

%% file: evaluation.tex
We implemented HAS-GPU based on the Kubernetes and OpenFaaS platform. A new custom resource definition (CRD) and operator, \textit{HASFunc}, was designed and implemented to manage serverless inference functions. We deployed the HAS-GPU system on a GPU cluster with 10 GPUs and nodes on LRZ Compute Cloud. Each node features an NVIDIA Tesla V100 GPU with 16GB device memory and an Intel(R) Xeon(R) Gold CPU @ 2.40GHz with 20 cores and 368GB RAM. 

We utilized deep learning applications from the standard MLPerf benchmark \cite{reddi2020mlperf} to create the serverless inference function benchmark for our experiments. As for workload, we exploited the practical application workloads from Microsoft Azure Trace \cite{zhang2021faster} and used Grafana k6 as the load generator. 

For the RaPP training and evaluation, we constructed an inference latency dataset based on all official deep learning models on PyTorch running under various batch sizes, SM partitions, and time quota configurations. The dataset contains 53400 data samples. We randomly selected 42720 samples as the training set, 5340 samples as the validation set, and 5340 samples as the test set.

\vspace{-2mm}
\subsection{Model Performance with Fine-grained Resource Allocation}
\vspace{-4mm}
\begin{figure}[htbp]
  \centering
  \begin{subfigure}[b]{0.46\textwidth}
    \includegraphics[width=\textwidth]{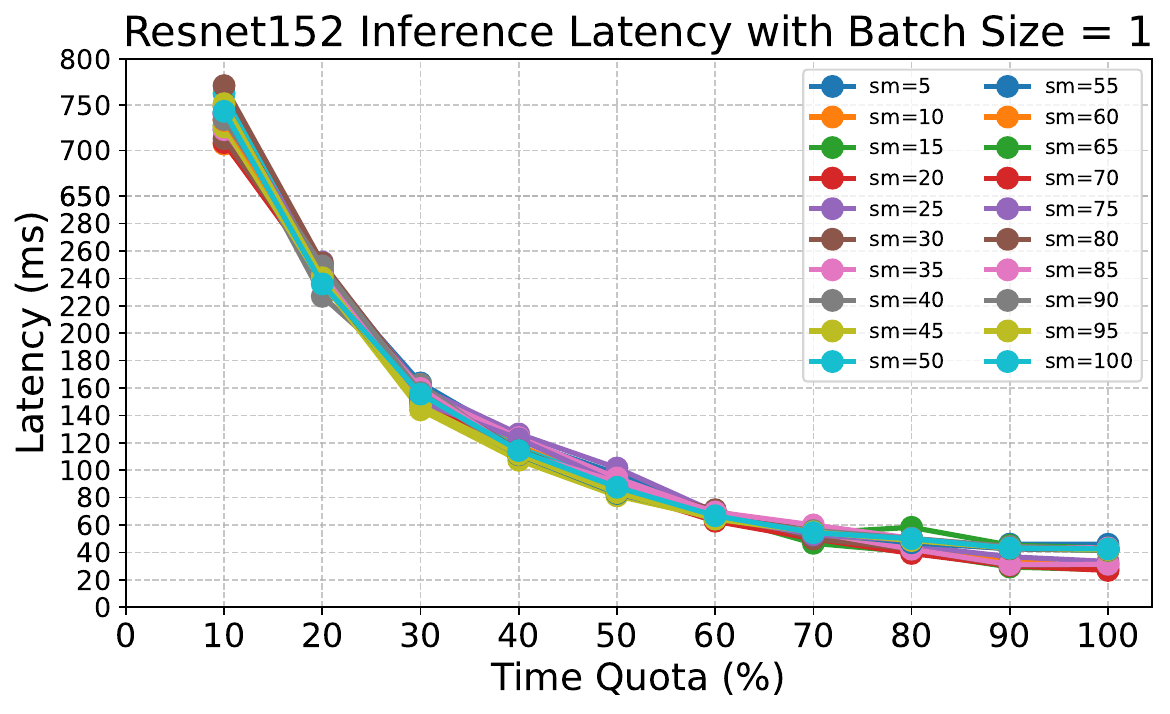}
  \end{subfigure}
  \begin{subfigure}[b]{0.46\textwidth}
    \includegraphics[width=\textwidth]{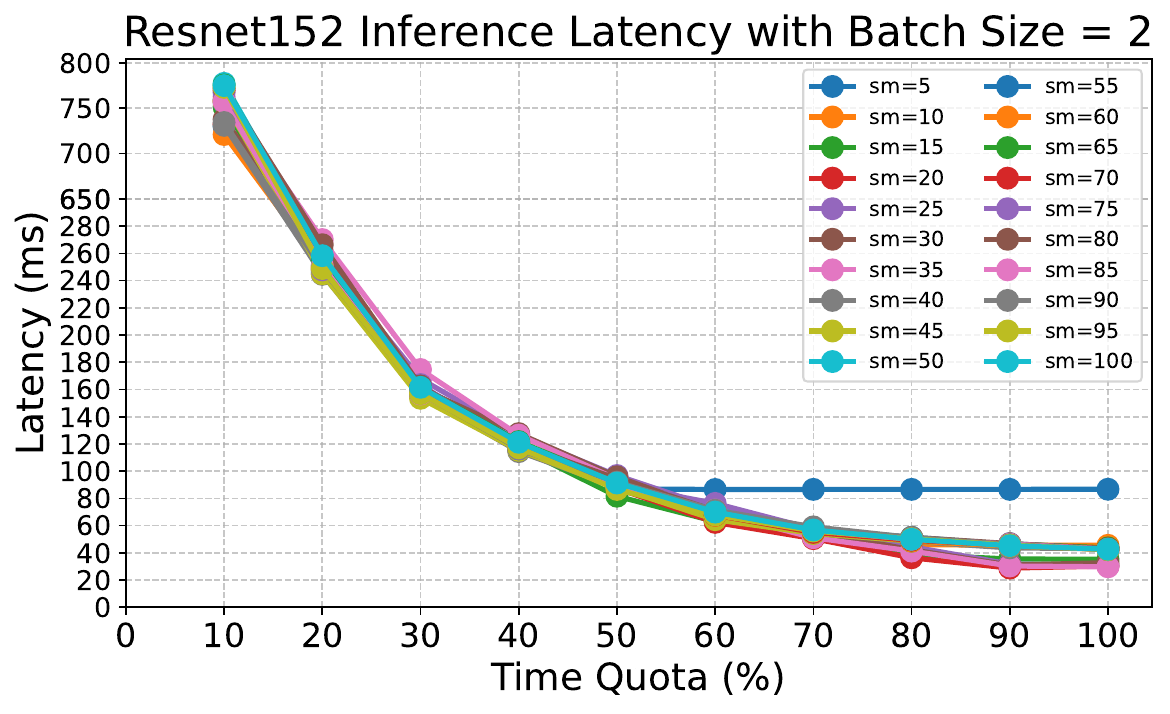}
  \end{subfigure}
  \begin{subfigure}[b]{0.46\textwidth}
    \includegraphics[width=\textwidth]{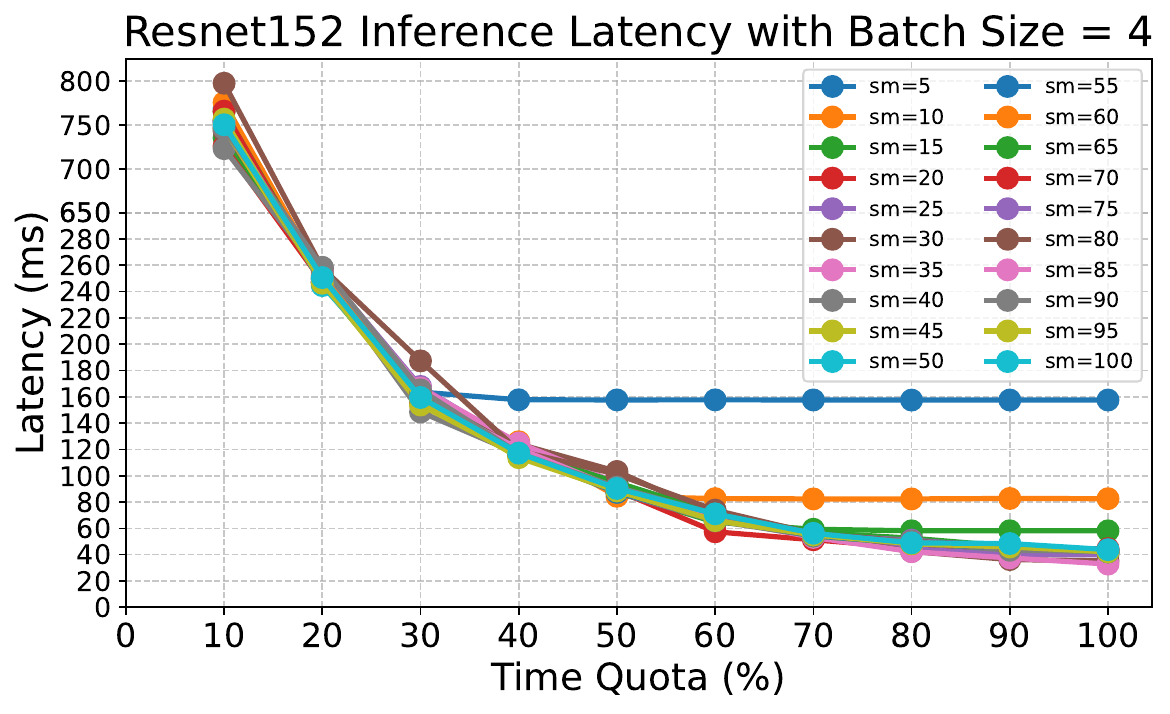}
  \end{subfigure}
  \begin{subfigure}[b]{0.46\textwidth}
    \includegraphics[width=\textwidth]{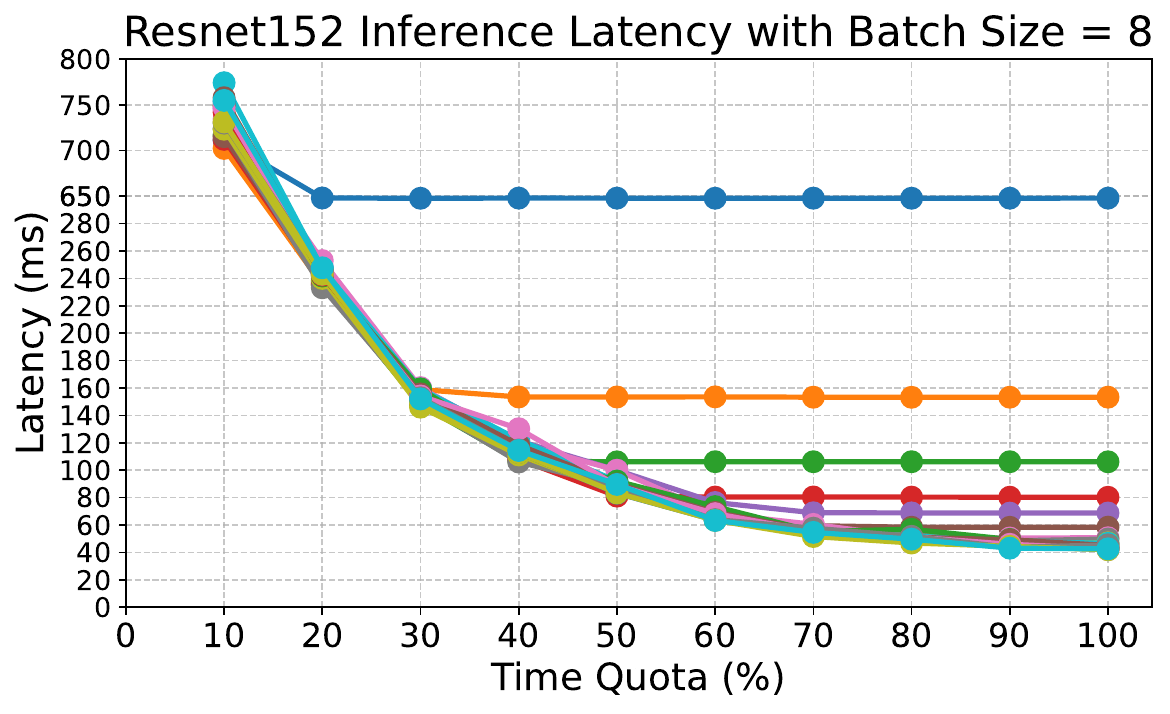}
  \end{subfigure}
  \vspace{-3mm}
  \caption{Inference latencies of Resnet152 under different configurations.}
  \label{fig:resnet152_latency}
  \vspace{-7mm}
\end{figure}
Figure \ref{fig:resnet152_latency} illustrates the inference latency of ResNet-152 under different batch sizes, SM partitions, and quota allocations. The results validate the effectiveness of HAS-GPU's fine-grained spatio-temporal resource allocation. With sufficient SM allocation, increasing the quota reduces inference latency and enhances throughput, demonstrating the effectiveness of quota reallocation-based vertical scaling. Since function throughput capability is defined as {\small{$\frac{\text{Batch}}{\text{Latency}}$}}, even minor latency reductions significantly boost throughput in low latency. Meanwhile, when the batch size is large and the SMs allocated to a function are insufficient, increasing the time quota does not reduce the latency. Conversely, for smaller batch sizes, allocating additional SMs also does not improve performance. These observations highlight the importance of performance prediction in fine-grained resource allocation.
\subsection{Resource-aware Performance Prediction Analysis}
\begin{figure}[htbp]
  \centering
  \begin{subfigure}[b]{0.45\textwidth}
    \includegraphics[width=\textwidth]{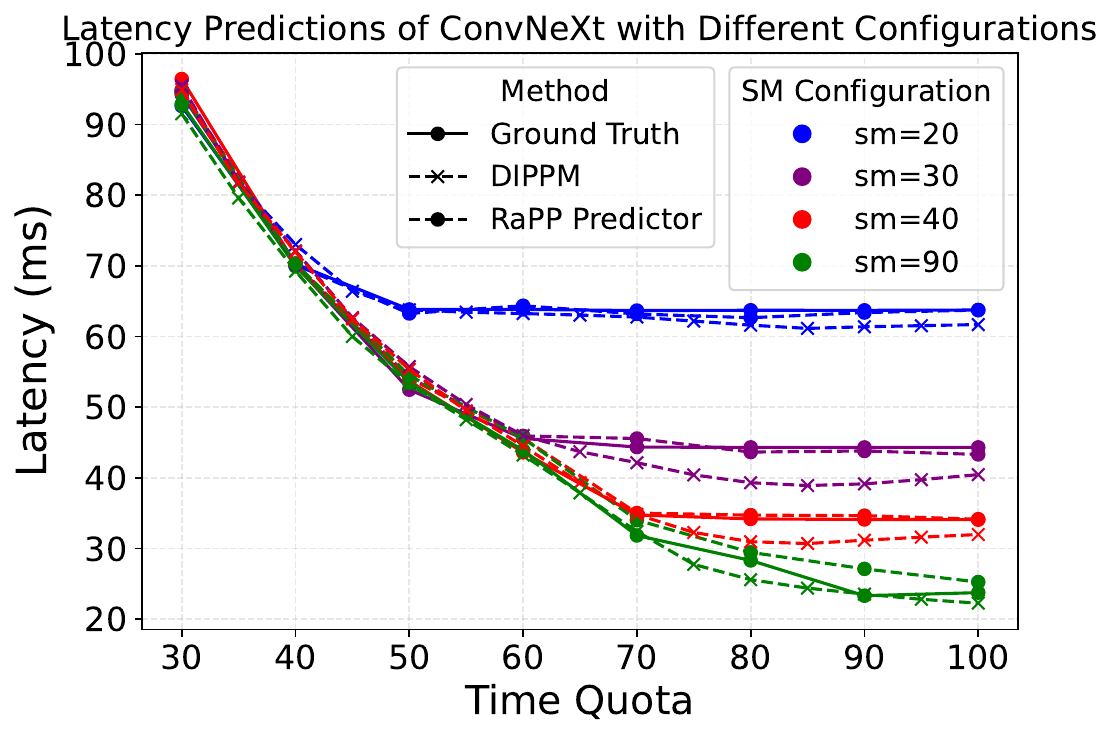}
  \end{subfigure}
  \begin{subfigure}[b]{0.4\textwidth}
    \includegraphics[width=\textwidth]{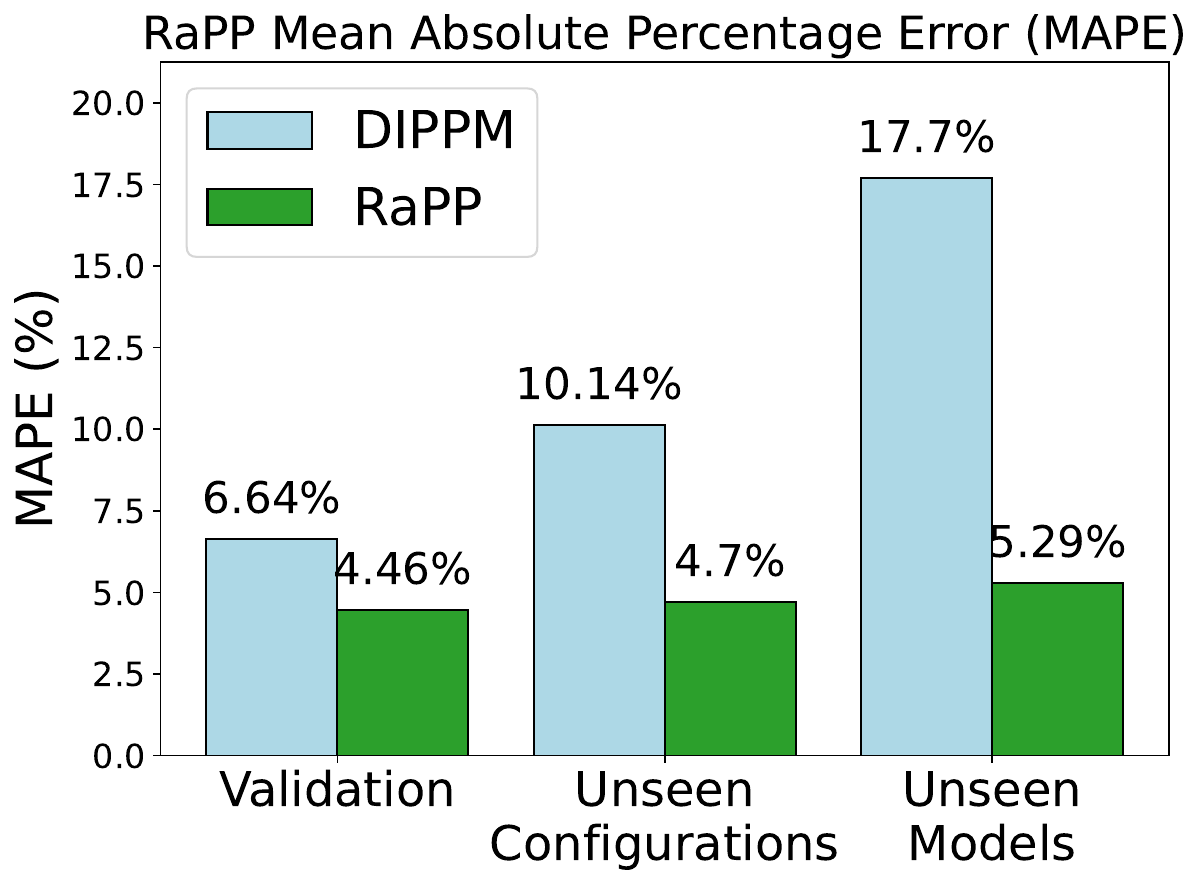}
  \end{subfigure}
  \vspace{-3mm}
  \caption{The latency prediction of the ConvNeXt and the accuracy of RaPP.}
  \label{fig:rapp_performance}
\end{figure}
Figure \ref{fig:rapp_performance} presents RaPP's latency predictions for the ConvNeXt model and the overall Mean Absolute Percentage Error (MAPE) of RaPP, compared against DIPPM \cite{panner2023dippm}, a method solely based on static model features. DIPPM does not support fine-grained resource configurations as input. For comparison, we incorporated this information into its static features same as RaPP and retrained the model. The result demonstrates that RaPP consistently aligns closely with the ground truth under various SM and quota resource allocations. RaPP maintains high prediction accuracy even for predicting small latency, whereas DIPPM shows significantly larger deviations. As for MAPE, RaPP maintains a latency prediction error of around 5\%, meaning a 20ms latency prediction deviates by less than 1ms.  It consistently outperforms DIPPM on both the validation and test sets, particularly for unseen configurations and models. While DIPPM's error rate rises from 10.14\% to 17.7\%, RaPP sustains a low error rate. This highlights the importance of extracting operator and graph runtime features, which enables RaPP to robustly adapt to fine-grained resource allocations.

\subsection{SLO Violation and Function Cost Analysis}
\begin{figure}[htbp]
  \centering
  \begin{subfigure}[b]{0.57\textwidth}
    \includegraphics[width=\textwidth]{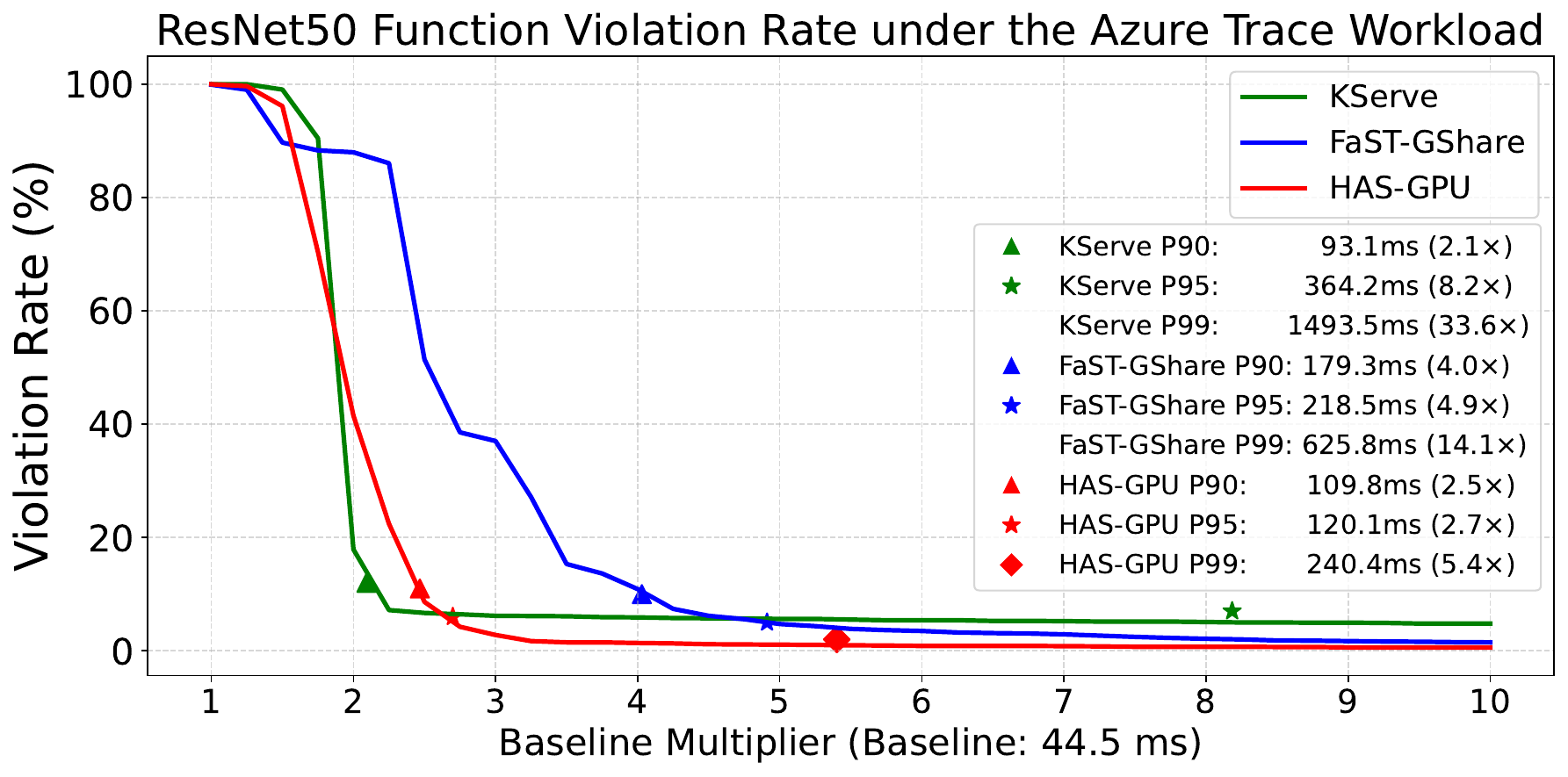}
  \end{subfigure}
  \begin{subfigure}[b]{0.40\textwidth}
    \includegraphics[width=\textwidth]{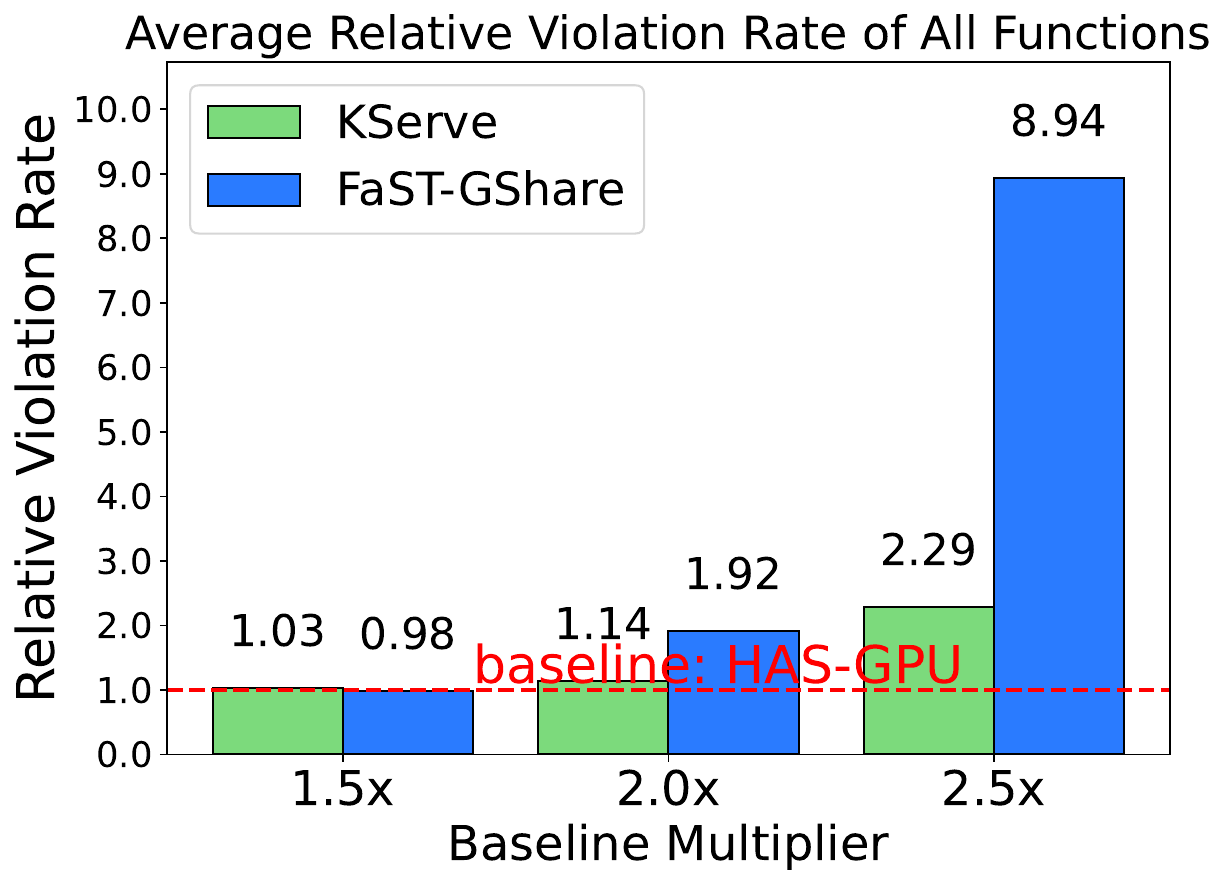}
  \end{subfigure}
  \vspace{-3mm}
  \caption{Function violation rates of ResNet50 and relative rates of all functions.}
  \label{fig:slo_violatioon}
\end{figure}
\vspace{-1mm}
To comprehensively reflect the function violations, we use the theoretical shortest inference time of a DL model running in a pure container as the baseline. With a step size of 0.25, we analyze the variation in function violation rate under baseline multipliers ranging from 1 to 10. Figure \ref{fig:slo_violatioon} shows the result of ResNet50 and relative violation rates of all benchmark functions with HAS-GPU's violation rate as the baseline. We compared the HAS-GPU system with the mainstream GPU serverless inference platform KServe \cite{kserve} and the state-of-the-art spatio-temporal GPU Sharing FaaS framework FaST-GShare \cite{gu2023fast}. Results from ResNet50 indicate that both HAS-GPU and KServe effectively reduce violation rates under smaller SLOs, while FaST-GShare maintains a higher violation rate. This is because HAS-GPU can quickly adapt to dynamic serverless workloads through vertical scaling, and KServe, with exclusive GPU allocation, benefits from higher concurrent processing capacity. In contrast, FaST-GShare relies on fixed fine-grained resource allocation and can only meet workload changes through horizontal scaling, where cold start delays contribute to its persistently high violation rate. We further analyze the performance of each method on P90, P95, and P99 metrics. HAS-GPU maintains low latency across all metrics, while KServe experiences significant delays at P95 and P99. This is due to KServe’s GPU instance-based horizontal scaling, which incurs high latency from GPU device and system initialization, leading to pronounced tail latency effects. In contrast, HAS-GPU’s vertical scaling provides buffer time for horizontal scaling, demonstrating the high reliability of hybrid auto-scaling. For all functions, HAS-GPU achieves lower SLO violations than the other two methods under tighter SLOs (baseline multipliers, 1.5x, 2.0x, 2.5x). Compared to FaST-GShare, HAS-GPU reduces SLO violations by an average of 4.8x.

\begin{figure}[htbp]
  \centering
  \begin{subfigure}[b]{0.47\textwidth}
    \includegraphics[width=\textwidth]{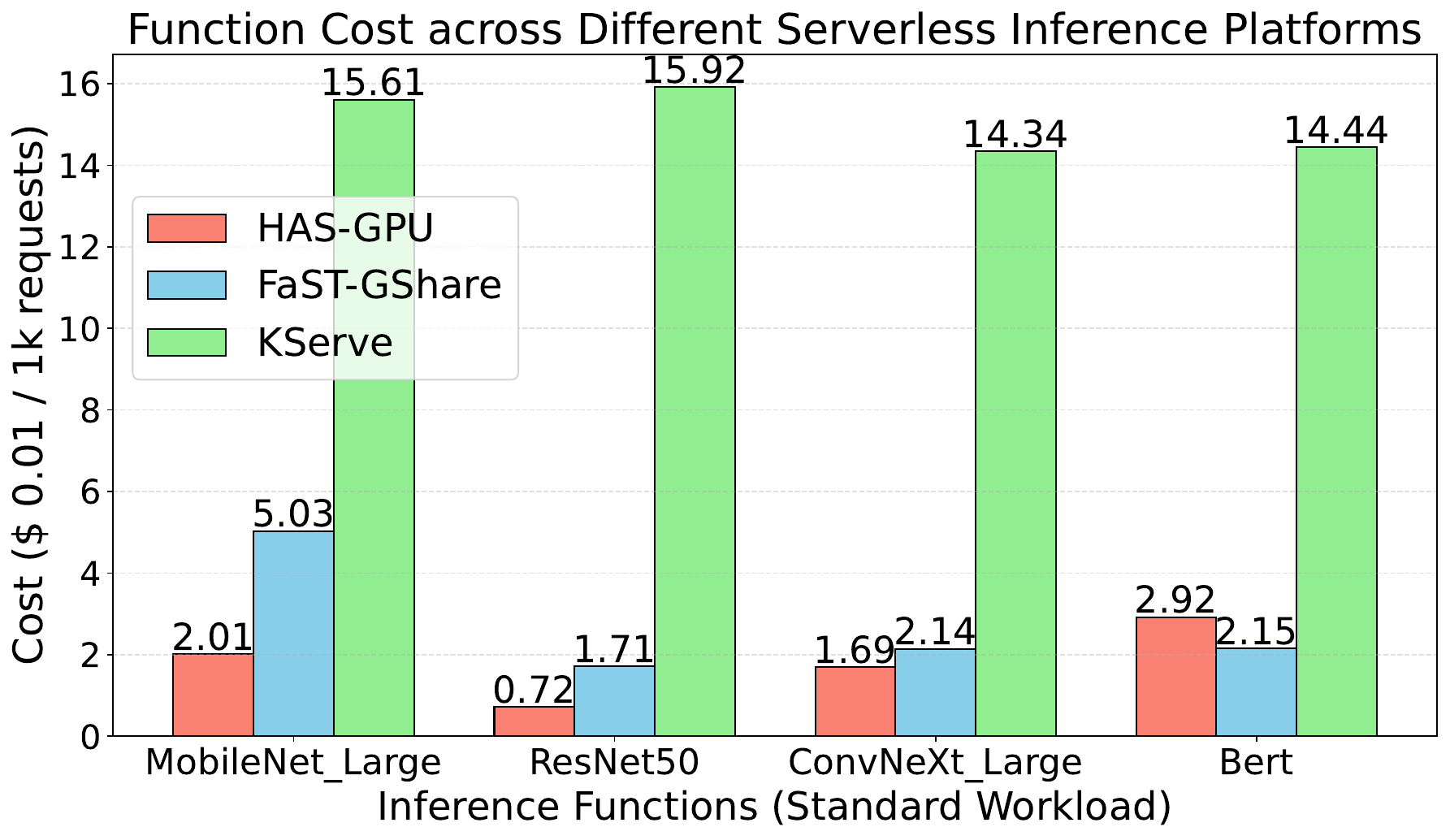}
  \end{subfigure}
  \begin{subfigure}[b]{0.47\textwidth}
    \includegraphics[width=\textwidth]{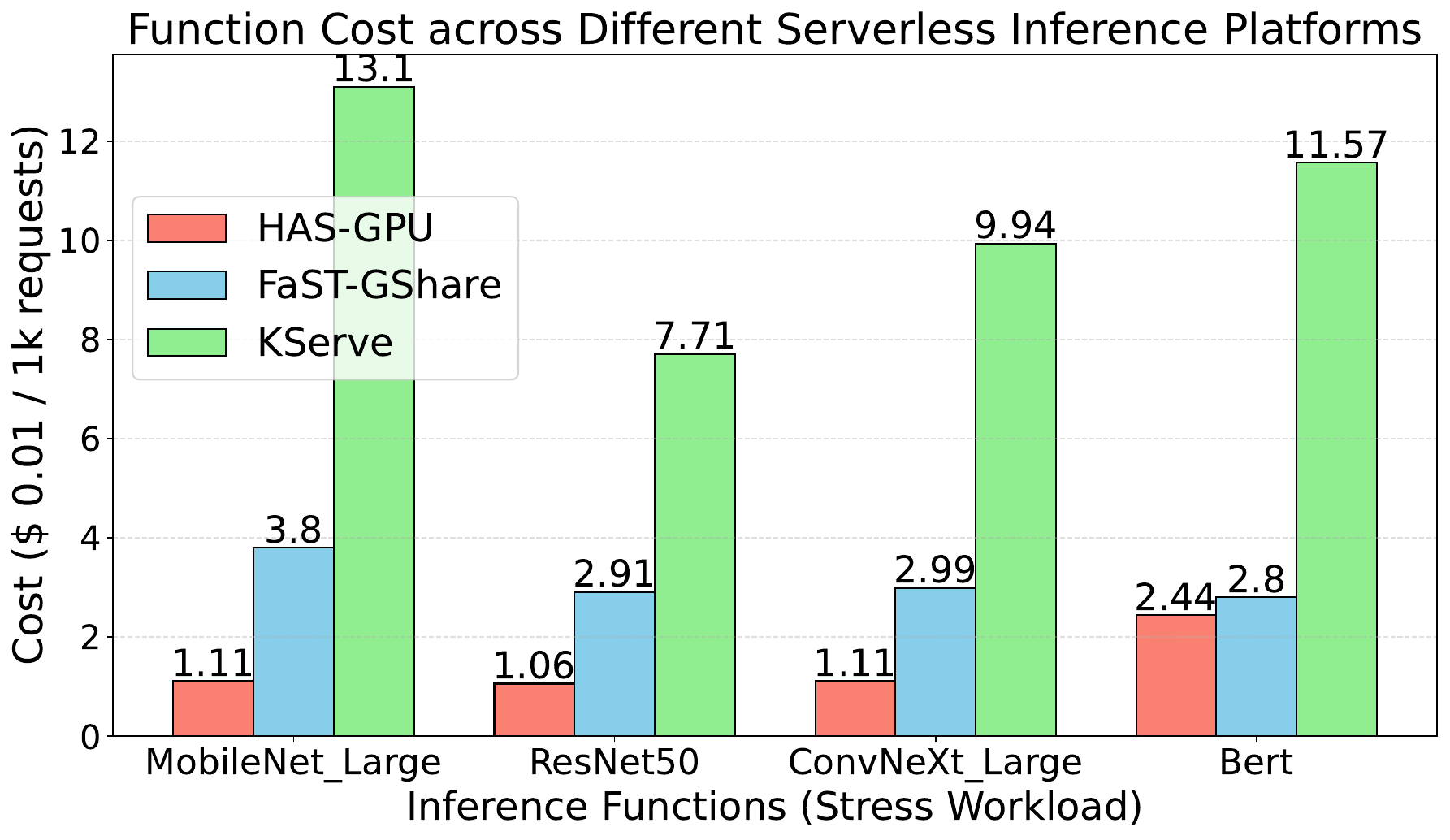}
  \end{subfigure}
  \vspace{-3mm}
  \caption{Function costs of different models under standard and stress workloads.}
  \label{fig:function_cost}
  \vspace{-7mm}
\end{figure}

Figure 7 illustrates the inference costs of each platform under standard and stress workloads. We calculate function costs based on the Google Cloud V100 GPU price (\$2.48/hour). For fine-grained GPU allocation, costs are measured using the actual GPU resources and time consumed per function. Since KServe exclusively occupies a GPU during scaling and frequently scales to handle fluctuating workloads, it incurs extremely higher costs per 1K requests. FaST-GShare, with its fixed resource allocation, lacks elasticity, making it more expensive than HAS-GPU. In contrast, HAS-GPU’s adaptive vertical scaling efficiently adjusts to workload variations, providing a significant cost advantage, especially under stress workloads. Under standard workloads, HAS-GPU reduces costs by up to 10.8x compared to KServe and 1.72x compared to FaST-GShare on average.

%% file: conclusion.tex
In this paper, we propose HAS-GPU, an efficient Hybrid Auto-scaling Serverless architecture with fine-grained GPU allocation for deep learning inferences. HAS-GPU proposes an agile scheduler capable of allocating SM partitions and time quotas with arbitrary granularity and enables significant vertical quota scalability at runtime. We propose the Resource-aware Performance Prediction model to address performance uncertainty introduced by massive configuration spaces. We present an adaptive hybrid auto-scaling algorithm to ensure inference SLOs and minimize GPU costs. The experiments demonstrated that, HAS-GPU reduces function costs by 10.8x on average compared to the mainstream serverless inference platform, and function SLO violations by 4.8x and cost by 1.72x compared to the state-of-the-art spatio-temporal GPU sharing FaaS framework.